# A New Ductile, Tougher Resin for Impregnation of Superconducting Magnets

**Emanuela Barzi[1,6], Daniele Turrioni[1], Ibrahim Kesgin[2], Masaki Takeuchi[3], Wang Xudong[4], Tatsushi Nakamoto[4], and Akihiro Kikuchi[5]**

[1] Fermi National Accelerator Laboratory, Batavia, IL 60510, USA
[2] Argonne National Laboratory, Lemont, IL 60439, USA
[3] RIMTEC Corporation, Kurashiki-city, Japan
[4] High Energy Accelerator Research Organization (KEK), Tsukuba, Ibaraki 305-0801, Japan
[5] National Institute for Materials Science, Tsukuba, Ibaraki 305-0047, Japan
[6] Ohio State University, Columbus, OH 43210, USA

E-mail: barzi@fnal.gov

**Abstract**

A major remaining challenge for $Nb_3Sn$ high field magnets is their training due to random temperature variations in the coils. The main objective of our research is to reduce or eliminate it by finding novel impregnation materials with respect to the epoxies currently used. An organic olefin-based thermosetting dicyclopentadiene (DCP) resin, $C_{10}H_{12}$, commercially available in Japan as TELENE® by RIMTEC, was used to impregnate a short $Nb_3Sn$ undulator coil developed by ANL and FNAL. This magnet reached short sample limit after only two quenches, compared with 50+ when CTD-101K® was used. Ductility, i.e. the ability to accept large strains, and toughness were identified as key properties to achieve these results. In addition, we have been investigating whether mixing TELENE with high heat capacity ceramic powders, increases the specific heat ($C_p$) of impregnated $Nb_3Sn$ superconducting magnets. The viscosity, heat capacity, thermal conductivity, and other physical properties of TELENE with high-$C_p$ powder fillers were measured in this study as a function of temperature and magnetic field. Mixing TELENE with either $Gd_2O_3$, $Gd_2O_2S$, and $HoCu_2$ increases its $C_p$ tenfold. We have also investigated the effect on the mechanical properties of pure and mixed TELENE under 10 Gy+ of gamma-ray irradiation at the Takasaki Advanced Radiation Research Institute in Takasaki, Japan. Whereas both TELENE-82wt% $Gd_2O_3$ and TELENE-83wt% $HoCu_2$ performed well, the best mechanical properties after irradiation were obtained for TELENE-87wt%$Gd_2O_2S$. Testing a short undulator in the future with the latter impregnation material will verify whether it will further improve the coils' thermal stability. Short magnet training will lead to better magnet reliability, lower risk and substantial saving in accelerators' commissioning costs.

Part of this study is supported by the U.S.-Japan Science and Technology Cooperation Program in High Energy Physics operated by MEXT in Japan and DOE in the U.S.

## 1. Introduction

One of the main challenges of Nb$_3$Sn high field accelerator magnets for HEP is their training [1]. Superconducting (SC) magnets go back to being resistive from their superconducting state, i.e. "quench", when their temperature increases above the current sharing temperature of the composite superconductor over a large enough volume. The temperature increase $\Delta T$ is proportional to $Q/C_p$, where Q is the dissipated heat, and $C_p$ is the volumetric heat capacity. Energy deposition that initiates quenches can emanate from a variety of both mechanical and electromagnetic sources (magnetic flux jumps, conductor motion, epoxy cracking, etc.). Other sources of magnet training are material interfaces, such as between conductor, insulation, impregnating material, and neighboring structural materials. All these sources contribute to a resulting "disturbance spectrum".

Long training has been a feature of any Nb$_3$Sn impregnated magnet for decades, since the start of the development of this technology, and any attempt made during this time to at least reduce magnet training failed. We show here almost total training elimination when using as coil impregnation material for a Nb$_3$Sn magnet $C_{10}H_{12}$, an organic olefin-based thermosetting dicyclopentadiene (DCP) resin, in replacement of the CTD-101K® epoxy currently used for this purpose. This resin is commercially available as TELENE® by RIMTEC Corporation, Japan, and its molecular structure is shown in Fig. 1, left. It was used to impregnate a Nb$_3$Sn short undulator model, which reached short sample limit (SSL) after only two quenches, compared with 50+ when CTD-101K (Fig. 1, right) was used on a number of identical undulator coils. TELENE's pot life of up to 3.5 hours at 5°C ensures scalability to impregnate larger coil volumes. The undulator magnet with 9 racetrack coils between 10 poles was wound at ANL [2]. After the winding was complete, the magnet was assembled into its reaction tooling to be heat treated in argon at FNAL using well established treatment cycles. It was then vacuum impregnated with pure TELENE at ANL, and later tested at FNAL in the Superconducting R&D lab [3].

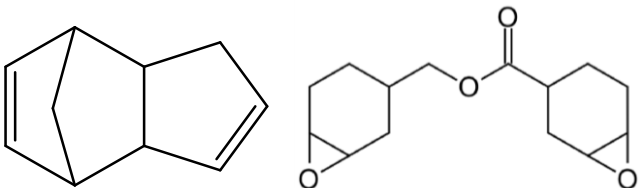

**Fig. 1.** Schematic of TELENE®, or $C_{10}H_{12}$, molecular structure (left); and of CTD-101 K epoxy, or $C_{21}H_{25}ClO_5$ (right).

To further improve thermal stability and training in accelerator magnets, the idea of increasing superconductor's stability, usually based on its Minimum Quench Energy (MQE), by inserting high specific heat (high-$C_p$) elements in superconducting wires dates back to the 1960s [4]. Then in the mid-2000s, a considerable improvement in stability to pulsed disturbances was obtained for NbTi windings, when distributing large heat capacity substances on the conductor during winding [5, 6]. The MQEs of the brushed coils were several times higher, and thermal efficiency was greatest for temperature diffusion times much smaller than the disturbance pulse duration. A few years ago, Hypertech and Bruker-OST have attempted to introduce high-$C_p$ elements in their wire design [7]. More recently, Hypertech fabricated samples of a thin composite Cu/Gd$_2$O$_3$ tape, which can be inserted in Rutherford-type cables to increase the conductor $C_p$ [8]. At NIMS and RIMTEC, TELENE was mixed with high-$C_p$ ceramic powders such as Gd$_2$O$_3$, Gd$_2$O$_2$S and HoCu$_2$ [9]. The $C_p$ temperature dependence for TELENE mixed with HoCu$_2$, and the $C_p$ temperature dependence as a function of magnetic fields were measured for TELENE mixed with Gd$_2$O$_3$, and Gd$_2$O$_2$S. NbTi superconducting wire samples impregnated with these resins were characterized and studied at FNAL by performing Minimum Quench Energy measurements.

The radiation strength of insulating materials used in superconducting accelerator magnets is another key specification. The common limit of the Hi-Lumi LHC type magnets is 25 MGy of proton radiation for CTD-101K epoxy. In 2016 the resistance to Cobalt-60 gamma radiation was studied for DCP and epoxy resin bisphenol-A up to a dose of 3.3 MGy with a dose rate of 2 kGy/h [10]. By measuring and analyzing optical absorption, electrical conduction, dielectric and thermal properties, it was shown that the organic DCP resin had a superior gamma ray resistance with respect to the epoxy. For nonorganic materials, there is a dependence of material response on the type of beam irradiation. However, such a dependence is quite modest for organic materials, and the absorbed dose can be adequately used to qualify their radiation resistance. Therefore, resistance to gamma irradiation is a promising indicator to radiation strength and a Cobalt-60 gamma ray irradiation experiment is being run at an average dose rate of 8 kGy/h at the Takasaki Advanced Radiation Research Institute [11], which is part of the National Institutes for Quantum Science and Technology (QST) in Takasaki. Here we present results of mechanical properties of pure and mixed TELENE before and during irradiation up to about 10 MGy.

## 2. Experiment Description

In this Section, we will describe the experimental setups used for mixing the TELENE with high-$C_p$ ceramic powders at NIMS (2.1); for measuring the resins' physical and mechanical properties at NIMS and KEK (2.2); for measuring the MQE of NbTi wire samples impregnated with the resins at FNAL (2.3); for fabricating and testing $Nb_3Sn$ undulator short models impregnated with TELENE at ANL and FNAL (2.4); and for the Cobalt-60 gamma ray irradiation experiment at the Takasaki Advanced Radiation Research Institute (2.5).

## 2.1 Fabrication of high-$C_p$ resins and optimization of their composition

The high heat capacity resins are fabricated by combining a ceramics powder filler with TELENE using a planetary mixer. The TELENE is then mixed with a hardener or polymerization catalyst, which is a ruthenium complex, in 2/100 parts by wt. The curing time is controlled by the amount of phosphine derivatives as retardant. The viscosity of the resins is controlled by the volume fraction and average size of the powder filler. Therefore, the chemical composition, average powder size and volume fraction can be optimized. Using a $Gd_2O_3$ powder size of 0.7 to 1.2 μm, TELENE was mixed using three different concentrations, i.e. 45wt%, 61wt%, and 82wt%. Using a $Gd_2O_2S$ powder size of 10 μm, TELENE was mixed using seven different concentrations between 58wt% to 88wt%.

Because of its high-$C_p$ and also smaller mass attenuation coefficient than Gd for thermal neutrons, in 2021 NIMS fabricated the first $HoCu_2$ powder by gas atomization, obtaining a particle size of 80 μm. A particle size of less than 30 μm was eventually achieved using a standard melt and casting process followed by a first stage grinding with a jaw crusher machine, and a second stage finer grinding with a planetary mill machine. The produced powder (Fig. 2) was used as a filler for TELENE with an 83wt% concentration.

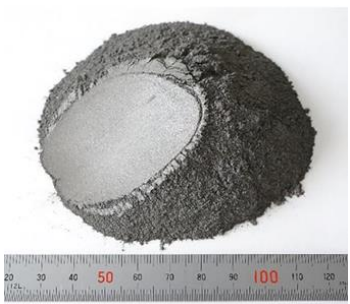

**Fig. 2.** Micrograph of the $HoCu_2$ particles produced by a standard melt and casting process.

## 2.2 Measurements of physical and mechanical properties of each resin

Physical properties of the resins, such as viscosity, thermal conductivity, and specific heat $C_p$ were measured at appropriate temperatures.

The viscosity was measured with a Brookfield-type viscometer, specifically an Eiko DV2T. Using a spindle speed of 60 rpm, a spindle of type LV-02 was used to measure viscosity values larger than 1 Pas, and one of type LV-04 to measure viscosity values lower than 1 Pas. The vscosity of TELENE was measured at 25°C, and that of CTD-101K at 60°C.

The $C_p$ and thermal conductivity were measured with a DynaCool® Physical Property Measurement System (PPMS) by Quantum Design. The $C_p$ temperature dependence for TELENE mixed with $HoCu_2$, and the $C_p$ temperature dependence as a function of magnetic field were measured for pure TELENE and TELENE mixed with $Gd_2O_3$, and $Gd_2O_2S$.

The mechanical properties that were measured for the resins include flexural modulus and flexural strength at room temperature.

The flexural modulus was obtained from the stress vs. strain curve between 0.05% strain and 0.25% strain.

The flexural strength was obtained through a 3-point bending test, of which a picture and schematic are shown in Fig. 3. These tests follow ISO 178:2010-A1:2013. Sample size is 80 mm in length, 10 mm in width and 4 mm in thickness. The flexural tests were performed at room temperature. The flexural strength is the maximum stress in the stress vs. strain curve. The tensile machine used were an Autograph AG-5000C manufactured by Shimadzu.

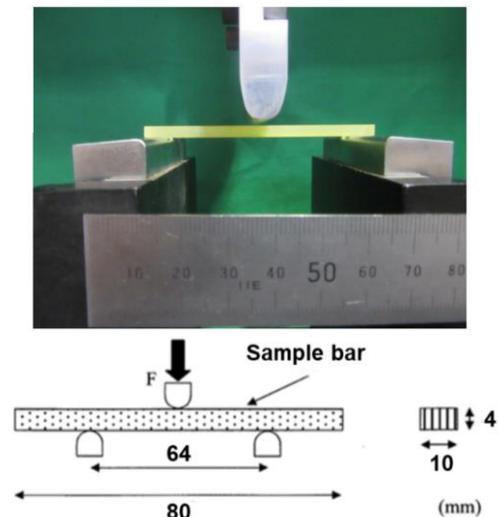

**Fig. 3.** Picture (top) and schematic of 3-point bending test (bottom).

## 2.3 Stability measurements of superconducting wire samples impregnated with high-$C_p$ resins

Two sets of six 0.8 mm NbTi wire samples were prepared at FNAL and sent to NIMS for impregnation with TELENE only, TELENE-82wt%$Gd_2O_3$, and TELENE-87wt%$Gd_2O_2S$ resins. The MQE of impregnated wires is measured on ITER-type barrels. Two strain gauges of 4 mm and 1.5 mm length and width are used as 350 Ω heaters and glued to each sample using STYCAST 2850FT. The instrumentation wires are soldered before sample and strain gauges receive resin impregnation. Fig. 4 shows pictures of instrumented NbTi wires after impregnation with TELENE and mixed TELENE.

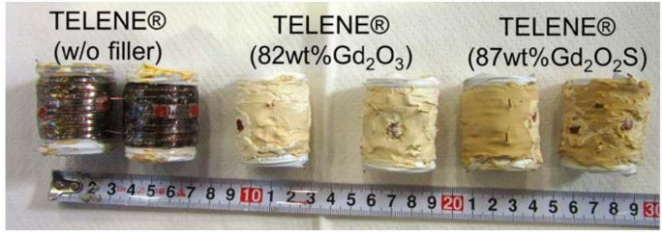

**Fig. 4.** Picture of TELENE impregnated NbTi wire samples instrumented with two heaters each, and with the voltage taps attached.

A 200 W power supply provides the excitation voltage to the strain gauges. Using a LabView DAQ program, a pulse output is generated from the power supply and the voltage across the strain gauge is measured. With the $I_c$ of the sample first measured, a constant bias current below $I_c$ is applied to the sample and heat pulses are fired using the strain gauge. A separate quench protection system monitors the voltage across the sample and shuts down the power supply if the quench threshold is reached. By gradually increasing the pulse energy, the minimum energy that induces a quench is defined as the MQE of the sample [8]. In order to determine the most appropriate pulse duration range for each resin, the characteristic time, or thermal time constant τ, was calculated as $τ = 4\ a^2/(π^2\ D)$, where $D = k/(ρ\ C_p)$ is the thermal diffusivity, k the thermal conductivity, ρ the material's density, and 2a the material's thickness. The thermal properties shown in Table 1 were obtained by using a = 1 mm, ρ(STYCAST 2850FT) = 2290 kg/m³, ρ(TELENE) = 1030 kg/m³, ρ(TELENE-82wt%Gd$_2$O$_3$) = 3504 kg/m³, and ρ(TELENE-87wt%Gd$_2$O$_2$S) = 4110 kg/m³. Based on the results for τ, the MQE was measured for heater pulse durations from 200 ms to 1.5 s.

TABLE 1: Thermal properties of TELENE resins

| @4.2 K | k W/(mK) | C$_p$ J/(kg K) | D m²/s | τ s |
|---|---|---|---|---|
| STYCAST 2850FT | 0.05 | 0.44 | 500.0 · 10⁻⁷ | 0.008 |
| TELENE | 0.04 | 3.5 | 111.0 · 10⁻⁷ | 0.037 |
| TELENE-82%Gd$_2$O$_3$ | 0.02 | 20 | 2.9 · 10⁻⁷ | 1.420 |
| TELENE-87%Gd$_2$O$_2$S | 0.09 | 60 | 3.7 · 10⁻⁷ | 1.111 |

## 2.4 Fabrication and testing of Nb$_3$Sn undulator short model

In collaboration with FNAL and other labs, ANL developed a Nb$_3$Sn undulator to be installed in the Advanced Photon Source (APS) storage ring. Performance reproducibility close to 100% short sample limit was obtained by using several Nb$_3$Sn short models during the R&D phase. They were vacuum impregnated with CTD-101K, which is the same epoxy used for Nb$_3$Sn high field accelerator magnets. The same performance and reproducibility were later achieved on longer models. The training behavior of the undulator models was very similar to that of High-Energy Physics accelerator magnets, sometime requiring almost 100 quenches to approach short sample limit [2].

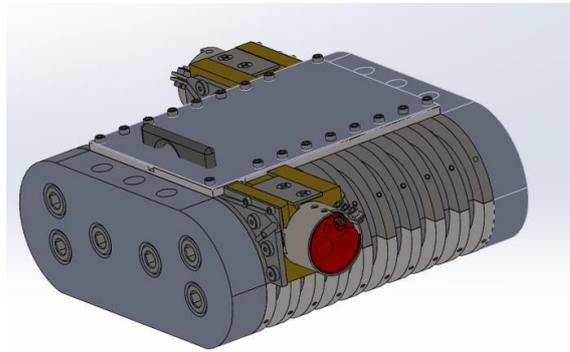

**Fig. 5.** 3D model of ANL undulator design.

These Nb$_3$Sn undulators with 18 mm period operate at a maximum magnetic field on the conductor of about 5 T. To address instabilities at this field, a Restacked Rod Processed (RRP) wire of 0.6 mm in diameter and with 144 superconducting subelements over 169 total subelements was used. Its equivalent subelement diameter is ~35 μm, and the critical current density $J_c$ (4.2K, 12 T) is about 2500 A/mm². Each Nb$_3$Sn undulator short model has nine racetrack coils wound in a groove between ten poles. There are 46 turns in each groove, and each period includes two grooves and two poles. The S2-glass braided Nb$_3$Sn wire is continuously wound turn-by-turn between the poles. A 3D model of the design is shown in Fig. 5.

After winding, the magnet was assembled into an existing reaction tooling. The magnet model was heat treated at FNAL in argon atmosphere in a 3-zone controlled tube furnace, using well-established treatment cycles [12]. Table 2 shows the nominal temperature values compared with the measured oven temperature. The temperature was averaged between two K-type calibrated and ungrounded thermocouples. Several witness samples of the same Nb$_3$Sn wire used in the coil were included in the furnace. Their critical current $I_c$ was determined from measuring the V-I curve using an electrical field criterion of 0.1 μV/cm. The calculation of the expected coil short sample limit is obtained by intersecting the average $I_c$ of these samples as function of the magnetic field with the magnet load line.

TABLE 2: Nominal vs. obtained heat treatment cycle for undulator short model impregnated with TELENE

| Nominal | | Obtained | |
|---|---|---|---|
| Time, h | T, °C | Time, h | $T_{Ave}$, °C |
| 48 | 210 | 48 | 207 |
| 104 | 370 | 104 | 365 |
| 50 | 650 | 50 | 647 |

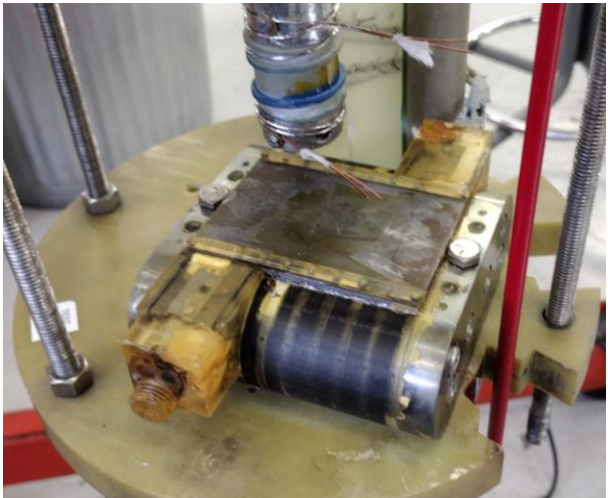

**Fig. 6.** Picture of the first TELENE impregnated Nb$_3$Sn small undulator attached to its test insert.

After winding and heat treatment, the magnet was vacuum impregnated at ANL. The undulator was later tested at FNAL at 4.2 K in liquid helium at atmospheric pressure, in a cryostat of the Superconducting R&D lab, by using an insert equipped with 2000 A DC leads. Fig. 6 shows the TELENE impregnated magnet attached to the test insert. Two pairs of voltage taps, each covering half of the magnet, were used. The voltage tap wires were connected to an NI-9239 card of a compact RIO DAQ system. The NI card has 4 channels with an acquisition frequency of 50 kHz and 24 bits per channel. The threshold for the quench protection system was 100 mV for the differential voltage. When a quench is detected, the power supply is stopped, an IGBT (insulated gate bipolar transistor) switch opens and the current flows into a 0.125 Ω dump resistor, where the coil energy gets dissipated.

*2.5 Gamma ray irradiation experiment*

Gamma Ray irradiation is being performed at the Takasaki Advanced Radiation Research Institute using a Cobalt-60 gamma irradiation facility. Thirtysix (36) samples each of pure TELENE, TELENE mixed with 82wt%Gd$_2$O$_3$, with 87wt%Gd$_2$O$_2$S, and with 83wt%HoCu$_2$ are being irradiated in air atmosphere at an average absorbed dose rate of 8 kGy/hr. Samples of CTD-101K epoxy were also included to verify the accuracy of the results. Fig. 7 shows the samples in their aluminum crate. The final goal for the entire irradiation campaign is to achieve 20 MGy+. Every month from start of irradiation, three samples of each resin were extracted from their aluminum rack and a 3-point bending test was performed at room temperature. Here we present results of mechanical properties of pure and mixed TELENE before and during irradiation up to about 10 MGy.

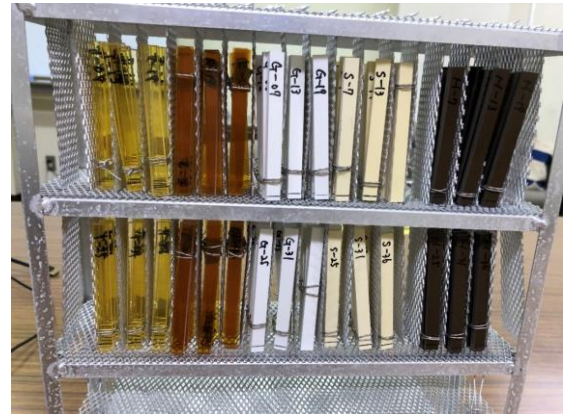

**Fig. 7.** Picture of aluminum crate containing the resins to be gamma ray irradiated in air atmosphere.

For nonorganic materials, there is a dependence of material response on the type of beam irradiation. However, such a dependence is modest for organic materials, and the absorbed dose can be used to qualify their radiation resistance. At a later stage, this could be confirmed with proton beam irradiation experiments at the BLIP facility at BNL.

**3. Results and Discussion**

TELENE was chosen for these studies because of the following main reasons: 1. Its ductility, i.e. the ability to accept large strains; 2. Its toughness, i.e. the amount of energy per unit volume that the material can absorb before rupturing, or the area underneath the stress vs. strain curve; 3. Its potential for radiation resistance. Fig. 8 shows how much more ductile and how much tougher is pure TELENE with respect to CTD-101K epoxy. Further, the potential of improving TELENE's thermal properties by mixing it with high-$C_p$ ceramic powders is a third strong component of interest for our study.

In this Section, we will present and discuss the results obtained for the resins' physical and mechanical properties (3.1.1 and 3.1.2); for the MQE of NbTi wire samples impregnated with the resins (3.2); for the TELENE impregnation and test of the first Nb$_3$Sn undulator short model, which includes impregnation process scalability (3.3.1), magnet short sample limits (3.3.2), and magnet test results (3.3.3); and for the Cobalt-60 gamma ray irradiation experiment up to 10 MGy at the Takasaki Advanced Radiation Research Institute (3.4).

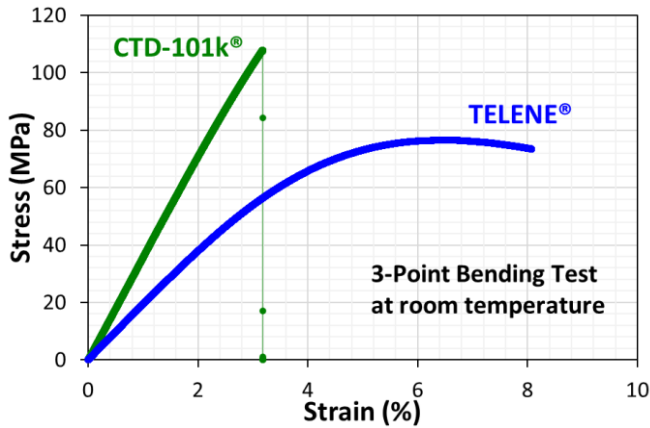

**Fig. 8.** Comparison of stress vs. strain curve between CTD-101K epoxy and TELENE resin at room temperature.

*3.1 Measurements of physical and mechanical properties of each resin*

*3.1.1 Physical properties.* Figs. 9 and 10 show respectively the thermal conductivity and specific heat as function of temperature for pure and mixed TELENE resins in absence of an external magnetic field. The specific heat as function of temperature at various external magnetic fields is shown in Fig. 11 for pure TELENE, in Fig. 12 for TELENE-45wt%$Gd_2O_3$, and in Fig. 14 for TELENE-87wt%$Gd_2O_2S$. The latter mixed resin has the largest thermal conductivity over the whole temperature range and a peak in $C_p$ between 4.3K and 5.3K at fields between 0 and 8 T. Pure TELENE has a $C_p$ which increases monotonically with temperature. Beyond 6K, the $C_p$ of pure TELENE is larger than that of TELENE mixed with $Gd_2O_3$ at any magnetic field.

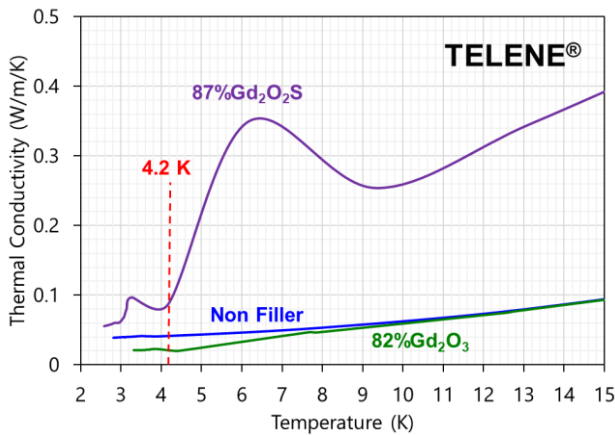

**Fig. 9.** Thermal conductivity vs. temperature for pure and mixed TELENE resins in absence of external magnetic field.

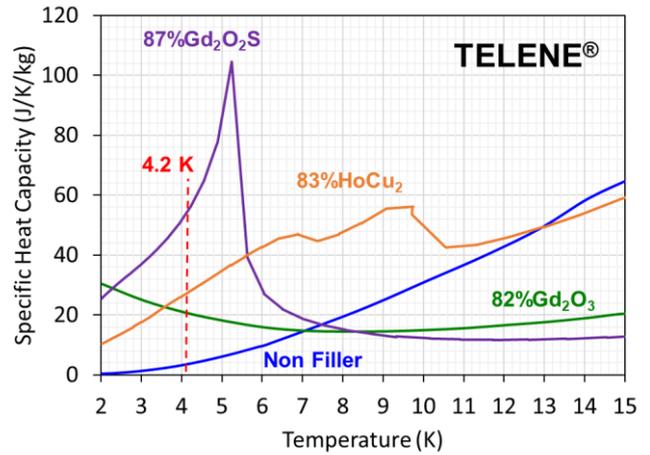

**Fig. 10.** Specific heat vs. temperature for pure and mixed TELENE resins in absence of external magnetic field.

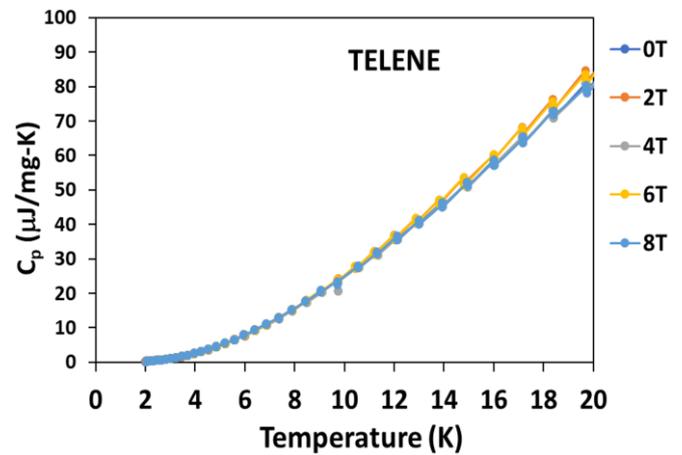

**Fig. 11.** Specific heat $C_p$ vs. temperature at various external magnetic fields for pure TELENE resin.

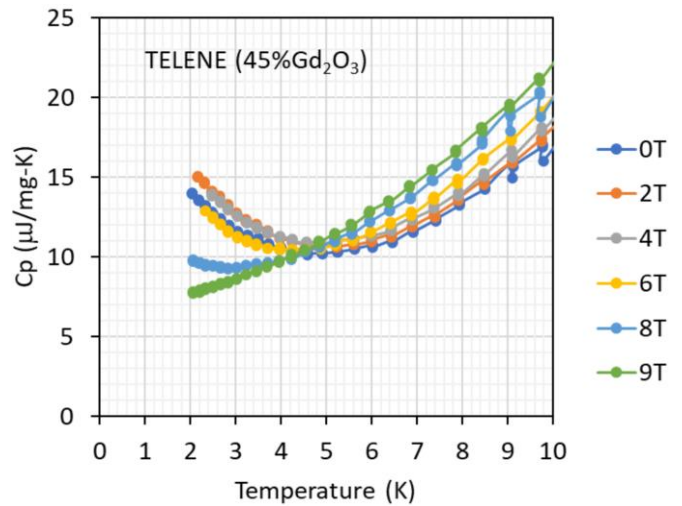

**Fig. 12.** Specific heat $C_p$ vs. temperature at various external magnetic fields for TELENE-45wt%$Gd_2O_3$ mixed resin.

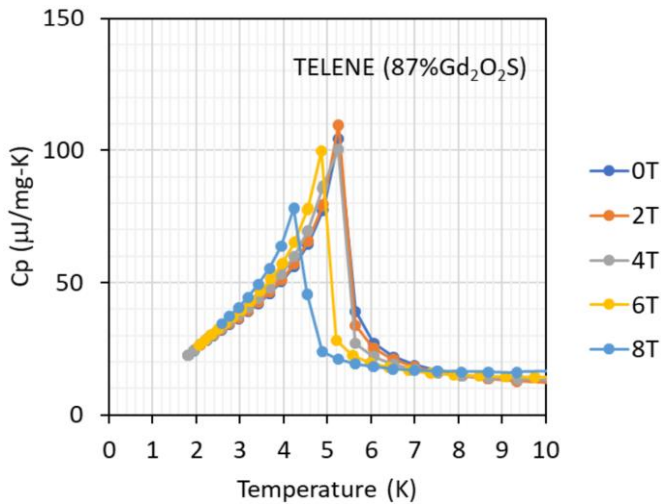

**Fig. 13.** Specific heat $C_p$ vs. temperature at various external magnetic fields for TELENE-87wt%$Gd_2O_2S$ mixed resin.

*3.1.2 Mechanical properties.* The flexural stress vs. strain curves for pure and mixed TELENE resins are shown in Fig. 14. After mixing the TELENE with hard ceramic particles, the material becomes stronger, i.e. larger flexural modulus, and less ductile. On the other hand, as seen in 3.1.1, some of the TELENE mixed resins feature larger thermal conductivity and specific heat than pure TELENE. It is reasonable to speculate that TELENE's ductility and capability to absorb large energies be key to the undulator training performance as detailed in Section 3.3. In a second part of this study, one has to check the impact of better thermal properties when using less ductile impregnation materials such as the high-$C_p$ mixed resins.

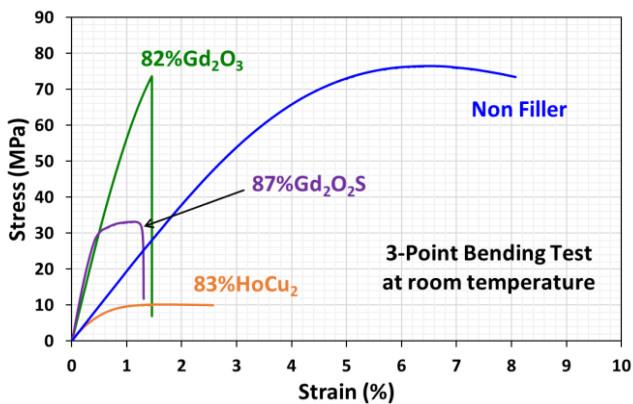

**Fig. 14.** Flexural stress vs. strain curves for pure and mixed TELENE resins.

## 3.2 MQE measurements of NbTi wire samples impregnated with high-$C_p$ resins

Based on the low diffusivity values obtained for the mixed TELENE resins shown in Table 1, with a maximum time constant of 1.42 s for TELENE-82wt%$Gd_2O_3$, the MQE of the impregnated 0.8 mm NbTi wire samples was measured for heater pulse durations from 200 ms to 1.5 s, with $I_c$% of up to 90% and magnetic fields between 6 and 9 T. At 9 T, the $I_c$(4.2K) was 140 A. An example of results obtained at 9 T and at 80% of $I_c$ is in Fig. 15. For pulse durations comparable to their time constant, both TELENE-82wt%$Gd_2O_3$ and TELENE-87wt%$Gd_2O_sS$ show larger increases in MQE than pure TELENE.

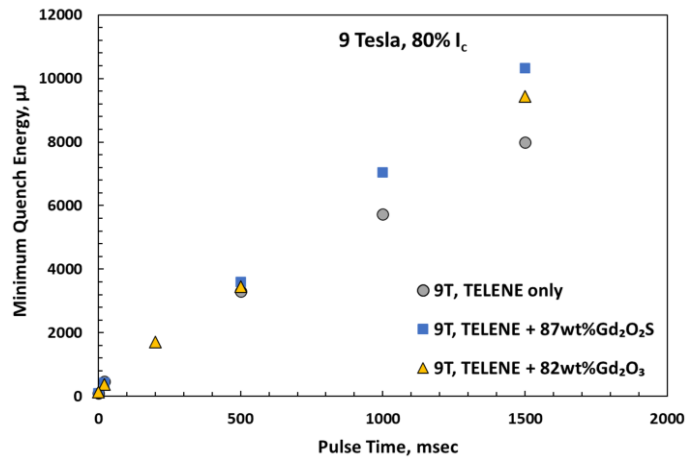

**Fig. 15.** Minimum Quench Energy vs. heater pulse duration at 80% of the critical current $I_c$ at 9 T for NbTi wire samples impregnated with pure and mixed TELENE.

## 3.3 Impregnation with TELENE and test of first $Nb_3Sn$ undulator short model

After winding and heat treatment, the magnet was placed in a leak-tight impregnation mold for vacuum impregnation at ANL. The two-part resin, i.e. TELENE resin plus the polymerization catalyst, was mixed by weight, and was then injected using positive pressure. After injecting the resin, the assembly was cured at 120°C for one hour. Due to the exothermic polymerization reaction, it is expected that the inside temperature is higher by about 50 to 100°C, depending on the amount of the resin.

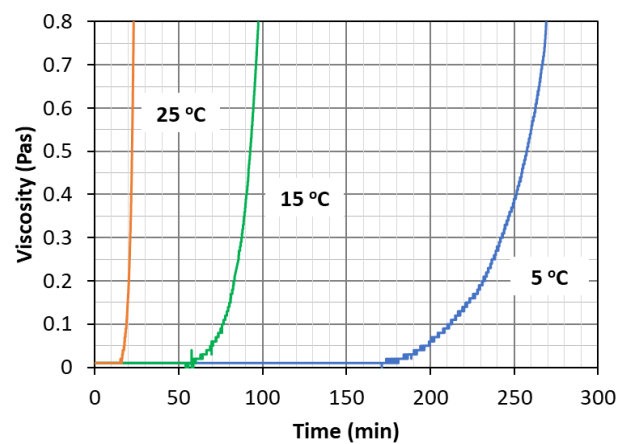

**Fig. 16.** Viscosity as function of time for TELENE at different temperatures.

As can be seen from Fig. 16, at room temperature, the pot life of TELENE is 20 minutes. However, the viscosity of TELENE is much lower than that of epoxy, i.e. its consistency is like water.

*3.3.1 Impregnation process scalability.* As shown in Fig. 17, TELENE's pot life can be increased by lowering the temperature during the impregnation process, which is the opposite of what is done for CTD-101 K, for which the temperature is instead increased. The dependence of TELENE's pot life with temperature is shown also in TABLE 3.

TABLE 3: Pot life vs. temperature for pure TELENE

| Pot life, min | Temperature,° C |
|---|---|
| 20 | 25 |
| 75 | 15 |
| 210 | 5 |

Scalability to larger impregnation volumes can be achieved by performing the impregnation process between 5 and 15°C. Indeed, by using one epoxy inlet into tooling equipped with multiple vents and an inlet pressure of 2 Bar, fill times with epoxy are less than 1.5 hrs for the HL-LHC IR quadrupoles that are 7.3 m long [13]. This includes about 45 minutes to inject CTD-101 K in the coil's mold and fill it, and about 40 minutes for filling the outflow tank.

The viscosity of mixed TELENE resins is of the same order of magnitude as that of pure TELENE up to high fillers concentrations, as shown for instance at 25°C for TELENE mixed with $Gd_2O_2S$ in Fig. 17. On the other hand, the viscosity of CTD-101K is much more sensitive to the amount of high-$C_p$ fillers, as shown for instance in Fig. 18 at 60°C when mixed with $Gd_2O_3$.

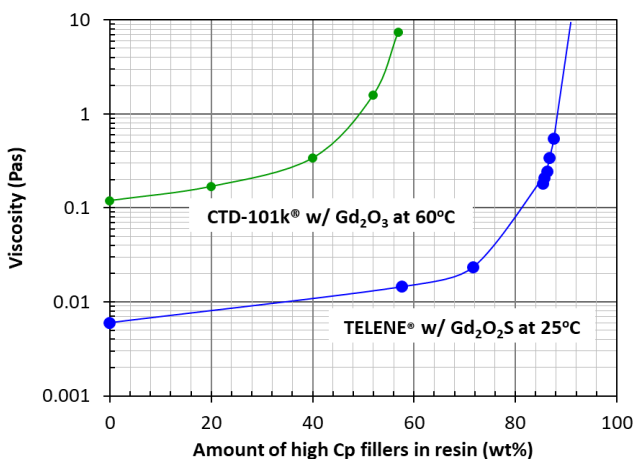

**Fig. 17.** Viscosity as function of wt%$Gd_2O_2S$ in TELENE at 25°C compared with that at 60°C of CTD-101K mixed with $Gd_2O_3$.

*3.3.2 Magnet short sample limits.* The short sample limit (SSL) for the first undulator short model was calculated based on the test results at 4.2K of three $Nb_3Sn$ witness samples that were included in the furnace with the coil. Fig. 18 shows that the $I_c$ vs. magnetic field curve for these samples intersects the maximum field load line of the undulator magnet at 1,143 A and 5.07 T.

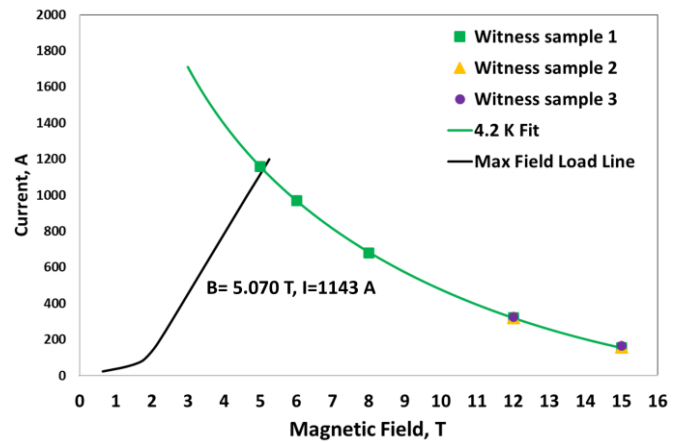

**Fig. 18.** Short sample limit calculation of TELENE impregnated undulator model based on witness sample critical current test results.

*3.3.3 Magnet test results.* The quench data for the first TELENE impregnated short undulator model as compared with identical undulator short models impregnated with CTD-101K are shown in Fig. 19. Data for model MM4 are not shown because it was damaged. Fig. 20 shows in more detail the quench history of this first undulator model, including a first thermal cycle. Actual quenches obtained at the standard ramp rate of 1 A/s are indicated with closed circles. Closed triangles indicate quenches produced during ramp rate studies, which were performed up to ramp rates of 40 A/s. Open circles represent errors or faulty trips.

As can be seen, the first quench at 1,043A occurred at about 91% of SSL. It took only three to four quenches to achieve 1,138 A, i.e. very close to the 1,143 A SSL predictions, compared to more than 50 quenches needed to reach a plateau for the nearly identical undulator coils impregnated with CTD-101 K.

The quench results during the first thermal cycle, i.e. second test sequence performed after warming up the magnet to room temperature and cooling it down again, is also shown in Fig. 20. In this second round, the first quench at 1,082A occurred at about 95% of SSL, and SSL was reached with just the second quench. However, this sequence also showed a number of current drops, from 3 to 8 % of SSL. The analysis of the voltage tap signals did not provide any insight on the nature of these quenches, and additional instrumentation will be needed to investigate this phenomenon.

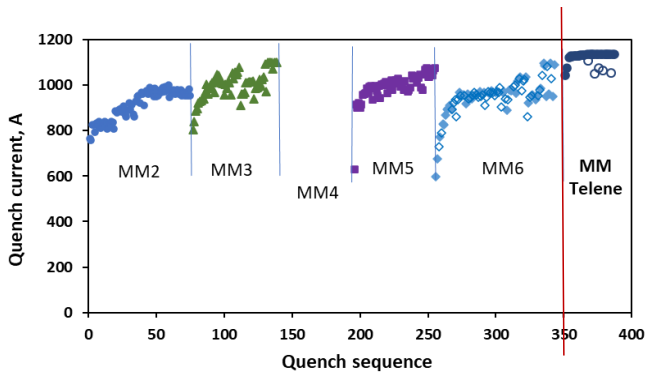

**Fig. 19.** Quench history of TELENE impregnated short undulator model as compared with that of nearly identical undulator short models impregnated with CTD-101K. Data for model MM4 are not shown because it was damaged.

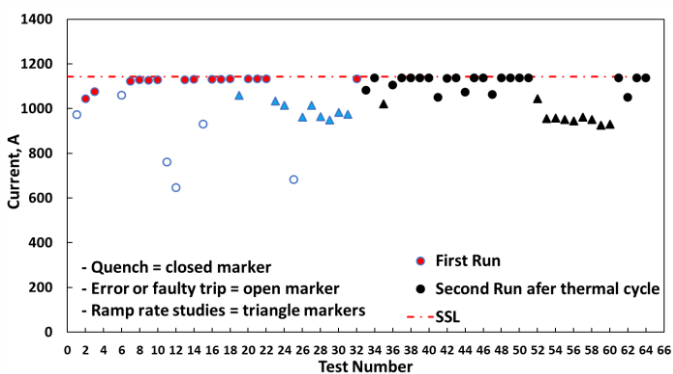

**Fig. 20.** Quench history, including first thermal cycle, for TELENE impregnated short undulator model. Actual quenches at the standard ramp rate of 1 A/s are indicated with closed circles. The maximum achieved current was 1,138 A.

*3.4 Cobalt-60 gamma ray irradiation experiment up to 10 MGy*

Figs. 21 and 22 show respectively the flexural strength and the flexural modulus as function of Gamma Ray dose for pure and mixed TELENE compared with CTD-101K. The flexural modulus monotonically increased by about 40% for pure TELENE, more than 60% for TELENE-82wt%$Gd_2O_3$ and TELENE-87wt%$Gd_2O_3$S, and more than 450% for TELENE-83wt%$HoCu_2$. The flexural strength monotonically increased for TELENE-82wt%$Gd_2O_3$ and TELENE-87wt%$Gd_2O_3$S.

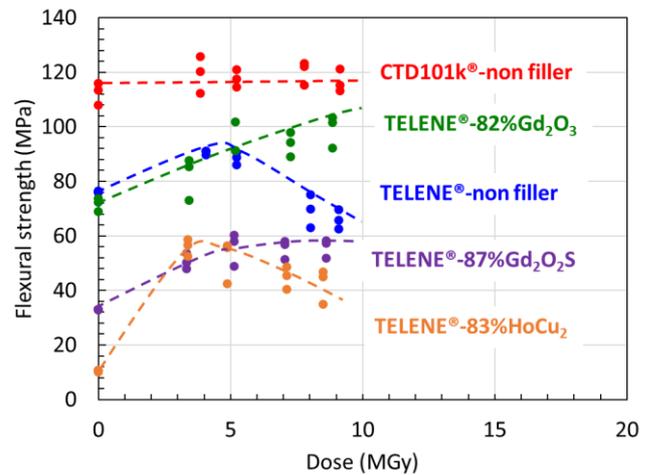

**Fig. 21.** Flexural strength as function of Gamma Ray dose for pure and mixed TELENE compared with CTD-101K.

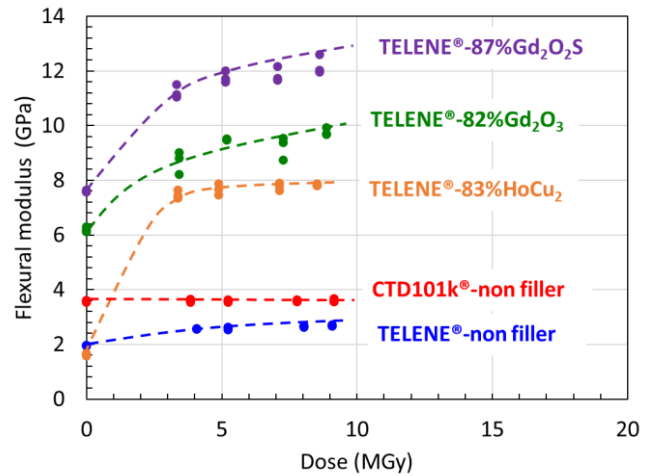

**Fig. 22.** Flexural modulus as function of Gamma Ray dose for pure and mixed TELENE compared with CTD-101K.

**4. Conclusions**

By using pure TELENE to impregnate a short $Nb_3Sn$ undulator coil developed with ANL, magnet training was nearly eliminated. This was attributed to TELENE's superior ductility and toughness.

Based on the higher $C_p$ and MQE results obtained for the mixed TELENE resins, two more ANL undulators are being wound for impregnation with TELENE-82wt%$Gd_2O_3$ and TELENE-87wt%$Gd_2O_2$S respectively to investigate any possible performance improvements produced by the better thermal properties of these mixed resins.

Gamma Ray irradiation performed at the Takasaki Advanced Radiation Research Institute using a Cobalt-60 gamma irradiation facility has shown that high-$C_p$ resins TELENE-82wt%$Gd_2O_3$ and TELENE-87wt%$Gd_2O_3$S are gamma ray resistant at least up to 10 MGy. This makes TELENE ideal for high radiation enviroments. The final goal for the entire irradiation campaign is to achieve 20 MGy+. At a later stage, TELENE's radiation resistance could be confirmed with proton beam irradiation experiments at the BLIP facility at BNL.

To study TELENE performance under larger Lorentz forces than those present in a 5 T light source undulator, pure TELENE and/or mixed TELENE will be used to impregnate LBNL Canted Cosine Theta sub-scale magnets, as well as FNAL High Temperature Supercondcuting dipole inserts, both developed within the U.S. Magnet Development Program.

To study TELENE performance under alternate loads in fast ramping magnets, such as those needed for a Muon Collider accelerator ring, pure TELENE and/or mixed TELENE will be used to impregnate FNAL Cosine Theta stressed managed coil made of superfine $Nb_3Sn$ wires, developed at NIMS for low AC losses, shaped in a multi-stage round cable.

By successfully reducing coil training, and based on the current radiation resistance results, TELENE impregnation technology is expected to have direct application to high field $Nb_3Sn$ dipole and quadrupole magnets, with substantial saving in accelerators commissioning costs.